\title{\huge Robust Bayesian hierarchical models for basket trials enabling joint evaluation of toxicity and efficacy}

\author{Zhi Cao$^{1\ast}$, Pavel Mozgunov$^{1}$, Haiyan Zheng$^{2}$ \\[8pt]
\textit{$^1$MRC Biostatistics Unit, University of Cambridge, U.K.} \\
\textit{$^2$Department of Mathematical Sciences, University of Bath, U.K.}
\\[8pt]
$^\ast$Email: \href{mailto:zhi.cao@mrc-bsu.cam.ac.uk}{zhi.cao@mrc-bsu.cam.ac.uk}
}

\date{ }

\documentclass[hidelinks, 11pt]{article}

\usepackage[margin=0.78in]{geometry}
\usepackage{hyperref}
\usepackage{amssymb, amsmath}
\usepackage{graphicx, caption}
\usepackage{booktabs}
\usepackage{multirow}
\usepackage{bm}
\usepackage{caption}
\usepackage{subcaption}

\usepackage{natbib}

\usepackage{pdfpages}

\usepackage{comment}

\begin{document}
\maketitle

\begin{abstract}
Basket trials have gained increasing attention for their efficiency, as multiple patient subgroups are evaluated simultaneously. Conducted basket trials focus primarily on establishing the early efficacy of a treatment, yet continued monitoring of toxicity is essential. In this paper, we propose two Bayesian hierarchical models that enable bivariate analyses of toxicity and efficacy, while accounting for heterogeneity present in the treatment effects across patient subgroups.  Specifically, one assumes the subgroup-specific toxicity and efficacy treatment effects, as a parameter vector, can be exchangeable or non-exchangeable; the other allows either the toxicity or efficacy parameters specific to the subgroups, to be exchangeable or non-exchangeable. The bivariate exchangeability and non-exchangeability distributions introduce a correlation parameter between treatment effects, while we stipulate a zero correlation when only toxicity or efficacy parameters are exchangeable. Simulation results show that our models perform robustly under different scenarios compared to the standard Bayesian hierarchical model and the stand-alone analyses, especially in producing higher power when the subgroup-specific effects are exchangeable in toxicity or efficacy only. When considerable correlation between the toxicity and efficacy effects exists, our methodology gives small error rates and greater power than alternatives that analyse toxicity and efficacy by parts.
\end{abstract}

\hspace{-1em} {\em Key words}: Bivariate analysis; Borrowing strength; Exchangeability; Phase II; Robustness

\section{Introduction} \label{sec:intro}

Precision medicine has been recognised as a new frontier for healthcare due to the rapid development of genetics and biomedical techniques \citep{Hodson2016, Akhoon2021}. Patient heterogeneity is a key consideration for tailoring treatments based on the disease features. Consequently, the paradigm of drug development gradually moves away from one-size-fits-all approaches \citep{marylene2015}. In recent decades, innovative trial methodologies such as {\em master protocols} have emerged as feasible and efficient approaches to drug development in the era of precision medicine.
Specifically, master protocols can streamline the design and analysis of clinical trials involving multiple patient strata to receive one or more treatments in a single overarching study \citep{park2019}.

Basket trials are one major type of master protocols in clinical trials: a new treatment can be evaluated simultaneously in several patient subgroups sharing a common disease feature which can, for example, be the same genetic aberration \citep{hobbs_basket_review}. For such commonality,  borrowing of information across patient subgroups is a preferred approach  to the analysis of basket trials. \citet{pf_bhm2003} discussed a Bayesian hierarchical model to estimate the efficacy parameters specific to subgroups under the assumption of exchangeability. Explicitly, subgroup-specific treatment effects are assumed as random samples from the same underlying distribution and the joint distribution remains invariant irrespective to their permutations \citep{Bernardo2001_exchangeability}. Such hierarchical structure is commonly implemented for information borrowing. However, assuming the exchangeability of all treatment effects can be overly restrictive, which can lead to inaccurate estimates for subtrials with extreme observations. Here, subtrials are formed by the patient subgroups each within a basket trial. To relieve the concern, \citet{exnex} proposed a robust extension that accounts for possibility that each of the subtrial-specific parameter may be non-exchangeable with the others. Other innovative approaches for robust borrowing of information have been under development these years. For example, \citet{hobbs} proposed exchangeability monitoring strategy using Bayesian monitoring rules for sharing information. \citet{hz_discrepancy} recommended information borrowing based on pair-wise discrepancies of treatment effects (monitoring the similarity of each pair of two subtrials to borrow information across those with high similarity). Besides, the adaptive clustering methods presuming more than one underlying distributions \citep{JackLee_BCHM} have been proposed to facilitate drug development.

Basket trials are often conducted in phase II oncology settings to evaluate the efficacy of an anti-cancer therapy, which targets specific molecular alterations across different tumour types. 
Oncologic basket trials are predominantly non-randomized that all patients within a subtrial are treated with the single evaluated treatment (so-called single-arm design) for rapid assessment of efficacy. 
Notable examples include the BRAF-V600 cancer study utilizing vemurafenib, which demonstrated its applicability as a targetable oncogene across multiple malignancies \citep{braf-v600}. The application of basket trials has extended beyond oncology to therapeutic areas such as neurodegenerative diseases. For instance, in neurodegenerative disorders research, basket trials have the potential to optimise resource use by testing various interventions within a unified framework \citep{Cummings2022}. Recent advancements have also introduced randomized basket trials and the use of continuous endpoints, such as the IMPACT II trial (\href{https://clinicaltrials.gov/study/NCT02152254}{NCT02152254}) which compares a targeted treatment against the standard therapy. An ongoing basket trial enrolls patients with Primary Biliary Cholangitis and participants with Parkinson's disease, in which a continuous cognitive test score was measured to evaluate the efficacy (\href{https://www.isrctn.com/ISRCTN15223158}{ISRCTN15223158}), thereby enhancing the rigour of therapeutic evaluations and biomarker development by incorporating randomisation procedure.\citep{Saad2017}

In drug development, phase II clinical trials answer the question of whether the new treatment can be recommended to the confirmatory phase III trial. 
There is growing recognition that solely evaluating efficacy 
in phase II trials may neglect other crucial aspects, such as safety, tolerability, and broader pharmacodynamic responses. This can lead to misjudgments about a drug's overall viability, especially when safety and other effects remain inadequately explored. Therefore, efficacy outcomes in Phase II trials may not always translate to Phase III success due to these overlooked factors. Besides, current statistical methodology to formally monitor toxicity data generally concentrates on single experimental treatment arm. \citet{ivanova2005} proposed a method to continuously monitor toxicity in phase II trials as early-stopping rules.
\citet{ray2011} developed a mechanism combining toxicity monitoring with Simon's two-stage design. 
More recently, a Bayesian multinomial model \citep{sam2021} was established to jointly analyse discrete toxicity and efficacy. However, there is limited investigation concerning a formal appraisal of toxicity effects in phase II basket trials.

In this article, we develop robust bivariate Bayesian analysis models for phase II basket trials to jointly evaluate toxicity and efficacy outcomes. The toxicity and efficacy effects will be formulated to allow a robust information borrowing across patient subgroups. 
Our proposed methods can be reviewed as the extensions of the exchangeable-nonexchangeable (EXNEX) model by \citet{exnex}: flexible prior mixture weights can be assigned to the exchangeability or non-exchangeability distributions of toxicity and efficacy parameters that imply different borrowing structures. Possible correlations between the toxicity and efficacy parameters are also taken into consideration in the bivariate analysis. 

The remainder of this article is structured as follows. In Section \ref{sec:methods}, we describe how to model the basket trial data for a joint evaluation of toxicity and efficacy treatment effects. Two Bayesian models are proposed with consideration of robust information borrowing and inference. Section \ref{sec:sim} presents a simulation study to assess the operating characteristics in comparisons with alternative methods. Finally, Section \ref{sec:discuss} will present a discussion on the proposed approaches.

\section{Methods} \label{sec:methods}

\begin{figure}[h]
	\begin{center}
		\includegraphics[width=.8\textwidth, height=0.32\textheight]{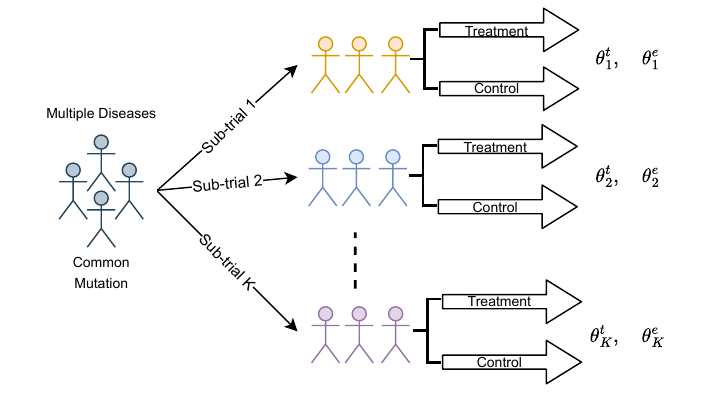}
	\end{center}
	\caption{Randomised controlled basket trial. Patients are divided into K subtrials, denoted by different colours. Notations: $\theta_k^t$ and $\theta_k^e$ used to denote the toxicity and efficacy-related effects specific to subtrial $k = 1, 2, \cdots, K$}
	\label{fig:our_basket_trial}
\end{figure}

Consider a phase II randomised controlled basket trial with $K$ subtrials, wherein patients are randomised to receive either the new treatment or control.
Each subtrial is characterized by a patient subgroup of specific disease subtype. Binary toxicity endpoint and continuous efficacy endpoint will be considered in this work. 
Figure \ref{fig:our_basket_trial} visualizes the configuration of such randomised controlled basket trials.

Let $n_{jk}$ be the sample size for subtrial $k=1,2,\cdots,K$, where $j = E, C$ labels the experimental or control group. We assume a binary toxicity outcome and a continuous efficacy outcome, as is commonplace in oncology settings with the RECIST guidelines \citep{Recist_guidelines2009}. Efficacy biomarkers include blood pressure, sample percentiles, measures of depression, etc.

\subsection{Data models for toxicity and efficacy outcomes} \label{sec:model_outcomes}


Let $y_{jk}$ be the number of toxicity responses in treatment group $j$ within subtrial $k$, which follows a $\text{Binomial}(n_{jk}, p_{jk})$ distribution with the response probability of $p_{jk}$.
We employ the binomial model with the ``logit" link function: for $k=1,\cdots,K$, 
\begin{equation}
\begin{aligned}
	&y_{jk} \sim \text{Binomial}\left( n_{jk}, p_{jk} \right) \\
	&\text{logit}(p_{jk}) = \alpha_{k}^{t} + T(j)\cdot \theta_k^t
\end{aligned}
\end{equation}
where $T(j)$ is the indicator function which equals 1 if $j=E$ or 0 if $j = C$. After the logit transformation, we can model the unknown baseline effect $\alpha_{k}^t$ (superscript $t$ for toxicity indication), and the toxicity treatment effect $\theta_{k}^t$.

Assuming normally distributed data, let $z_{ijk}$ be the efficacy outcome for patient $i$ receiving treatment $j$, within subtrial $k$. A patient-level efficacy model can be established as
\begin{equation}
\begin{aligned}
	&z_{ijk} \sim \mathcal{N} (\mu_{jk}, \sigma_{jk}^2)\\
	&\mu_{jk} = \alpha_{k}^e + T(j)\cdot \theta_k^e
\end{aligned}
\end{equation}
Unlike the toxicity model, the identity link is chosen, giving rise to the equivalence between the baseline effect $\alpha_{k}^e$ (superscript $e$ for efficacy indication) and the mean responses for the control group $\mu_{Ck}$ in subtrial $k$. Efficacy treatment effect $\theta_k^e$ can be established similarly to the toxicity model.

\subsection{BiEXNEX model} \label{sec:biexnex}

Building upon the idea of exchangeable-nonexchangeable (EXNEX) model by \citet{exnex}, we propose a bivariate Bayesian hierarchical model allowing for robust information borrowing across patient subgroups to jointly analyse the toxicity and efficacy effects. We also consider possible correlations between the two effects in the modelling.

The first hierarchical model (Bivariate EXNEX, abbreviated to BiEXNEX hereafter) is established by assuming the subtrial-specific parameter vectors of toxicity and efficacy effects are either exchangeable or non-exchangeable. That is, for each subtrial $k$:

with prior probability $\omega_k$, parameter vectors are exchangeable (EX):
\begin{subequations} \label{eq:biexnex}
\begin{equation}
\left.
\begin{pmatrix}
	\theta_k^t \\
	\theta_k^e
\end{pmatrix}
\right|
\begin{pmatrix}
	\beta_1 \\
	\beta_2
\end{pmatrix},
\begin{pmatrix}
	\phi_1 \\
	\phi_2
\end{pmatrix}, \rho
\sim
\mathcal{N} \left( 
\begin{pmatrix}
	\beta_1 \\
	\beta_2
\end{pmatrix},
\begin{pmatrix}
	\phi_1^2 & \rho\phi_1\phi_2 \\
	\rho\phi_1\phi_2 & \phi_2^2
\end{pmatrix}
\right),
\end{equation}
\quad\text{ } with prior probability $1 - \omega_k$, parameter vectors are non-exchangeable (NEX):
\begin{equation}
\left.
\begin{pmatrix}
	\theta_k^t \\
	\theta_k^e
\end{pmatrix}
\right| \kappa\quad
\sim \quad
\mathcal{N} \left( 
\begin{pmatrix}
	m_{1k} \\
	m_{2k}
\end{pmatrix},
\begin{pmatrix}
	s_{1k}^2 & \kappa s_{1k} s_{2k} \\
	\kappa s_{1k} s_{2k} & s_{2k}^2
\end{pmatrix}
\right).
\end{equation}
\end{subequations}

\noindent This BiEXNEX model contains prior probability of exchangeability, $\omega_k\in (0, 1)$, for each subtrial $k$, which may be specified according to historical knowledge or advice from practitioners.
The unknown parameters $(\beta_1, \beta_2)^T$ and $(\phi_1, \phi_2)^T$ are the mean vector and standard deviations in the covariance matrix of the EX part shared by all subtrials, with $\rho$ denoting the unknown correlation following the uniform prior $\mathcal{U}(-1, 1)$. $\beta_1$ and $\beta_2$ can be inferred with a normally distributed prior $\mathcal{N}(0, b^2)$, where $b$ is the prior standard deviation. The general prior selection for $\beta$ can be uninformative or weakly informative according to prior knowledge and clinicians' perspectives. For uninformative priors, $b$ can be set to a very large value, ensuring the toxicity difference between control and treatment groups covers a broad range of values. Specifying priors for $\phi^2$ is critical in the Bayesian setting - it can be interpreted as the heterogeneity between patient subgroups. Half-normal, half-Cauchy and half-t distributions can be chosen as candidates \citep{Gelman_hm_prior, Nicholas_half_cauchy_prior, book_prior}. We assume it follows a half-normal distribution $\text{Half-}\mathcal{N}(x)$ where $x$ determines the prior degree of information borrowing across patient subgroups.

In the NEX part, the mean $(m_{1k}, m_{2k})^T$ and standard deviation $(s_{1k}, s_{2k})^T$ for subtrial $k$ are fixed and elicited before applying the method to trial data, except for the correlation $\kappa$ uniformly distributing between -1 and 1. The NEX distribution is sometimes introduced as the ``operational prior" which are calibrated to accommodate considerable variability in the toxicity and efficacy effects across subtrials.
Such operational priors are often vague or weakly informative.

The BiEXNEX model shows its simplicity as a bivariate analysis method with only two mixture components to estimate the subtrial-specific parameters. Yet it could be limited for circumstances when either toxicity or efficacy parameters only are exchangeable across subtrials, which we refer to as ``asynchronous  exchangeability". 

\subsection{E-BiEXNEX model} \label{sec:ebiexnex}

We relax the assumption of simultaneously exchangeable treatment effects, and propose a more flexible model to extend BiEXNEX method, hereafter referred to as E-BiEXNEX. The E-BiEXNEX model for subtrial $k$ is structured with four mixture components to accommodate asynchronous exchangeability, coupled with four sets of prior probabilities of exchangeability:

with probability $\lambda_{1k}$, both $\theta_{k}^t$ and $\theta_{k}^e$ are exchangeable across subtrials,
\begin{subequations}
\begin{equation}
\left.
\begin{pmatrix}
	\theta_k^t \\
	\theta_k^e
\end{pmatrix}
\right| , 
\begin{pmatrix}
	\beta_1 \\
	\beta_2
\end{pmatrix},
\begin{pmatrix}
	\phi_1 \\
	\phi_2
\end{pmatrix}, \rho
\sim
\mathcal{N} \left( 
\begin{pmatrix}
	\beta_1 \\
	\beta_2
\end{pmatrix},
\begin{pmatrix}
	\phi_1^2 & \rho\phi_1\phi_2 \\
	\rho\phi_1\phi_2 & \phi_2^2
\end{pmatrix}
\right)
\end{equation}

with probability $\lambda_{2k}$, only $\theta_k^t$ is exchangeable,
\begin{equation}
\left.
\begin{pmatrix}
	\theta_k^t \\
	\theta_k^e
\end{pmatrix}
\right|
\beta_1,
\phi_1
\sim
\mathcal{N} \left( 
\begin{pmatrix}
	\beta_1 \\
	m_{2k}
\end{pmatrix},
\begin{pmatrix}
	\phi_1^2 & 0 \\
	0 & s_{2k}^2
\end{pmatrix}
\right)
\end{equation}

with probability $\lambda_{3k}$, only $\theta_k^e$ is exchangeable,
\begin{equation}
\left.
\begin{pmatrix}
	\theta_k^t \\
	\theta_k^e
\end{pmatrix}
\right|
\beta_2,
\phi_2
\sim
\mathcal{N} \left( 
\begin{pmatrix}
	m_{1k} \\
	\beta_{2}
\end{pmatrix},
\begin{pmatrix}
	s_{1k}^2 & 0 \\
	0 & \phi_2^2
\end{pmatrix}
\right)
\end{equation}

with probability $\lambda_{4k}$, neither $\theta_k^t$ nor $\theta_k^e$ is exchangeable,
\begin{equation}
\left.
\begin{pmatrix}
	\theta_k^t \\
	\theta_k^e
\end{pmatrix}
\right| \kappa \quad
\sim \quad
\mathcal{N} \left( 
\begin{pmatrix}
	m_{1k} \\
	m_{2k}
\end{pmatrix},
\begin{pmatrix}
	s_{1k}^2 & \kappa s_{1k} s_{2k} \\
	\kappa s_{1k} s_{2k} & s_{2k}^2
\end{pmatrix}
\right)
\end{equation}
\end{subequations}
where $\Sigma_{i=1}^4\lambda_{ik} = 1$ and $0 \leqslant \lambda_{ik} \leqslant 1$ for each subtrial $k$. Similar to the parameter elaboration in the BiEXNEX model, $(\beta_1, \beta_2)^T$ and $(\phi_1, \phi_2)^T$ are unspecified and shared variables to allow for information borrowing among all subtrials, while $(m_{1k}, m_{2k})^T$ and $(s_{1k}, s_{2k})^T$ are fixed and precedingly specified for subtrial $k$. The prior specifications for (hyper)parameters in E-BiEXNEX model will stay the same as those in the BiEXNEX modelling.

The E-BiEXNEX model contains more configurations of the exchangeability and non-exchangeability distributions for toxicity and efficacy effects.
We thus expect this model to be more robust than the BiEXNEX model, which would facilitate the inferences in situations where the degrees of information borrowing differ across the toxicity and efficacy treatment effects. 
For E-BiEXNEX model, calculating posterior probabilities of ($\lambda_{1k} + \lambda_{2k}$) can approximate the posterior probability of exchangeability marginally for the toxicity effects, and likewise, posterior probabilities of ($\lambda_{1k} + \lambda_{3k}$) approximate that  for the efficacy effects. 
Supplementary material \ref{app:another_stipulation} provides proof of such approximation, for which we re-express the E-BiNEXNEX model
using ancillary categorical variables and conditional probabilities.

\subsection{Decision-making rule} \label{sec:dmrule}

On the completion of a basket trial, data are analysed to yield Go or No-go recommendations towards a confirmatory phase III study. Two decision criteria are elaborated below for joint evaluation of efficacy and toxicity.

\begin{figure}[ht]
	\begin{center}
		\includegraphics[width=.8\textwidth]{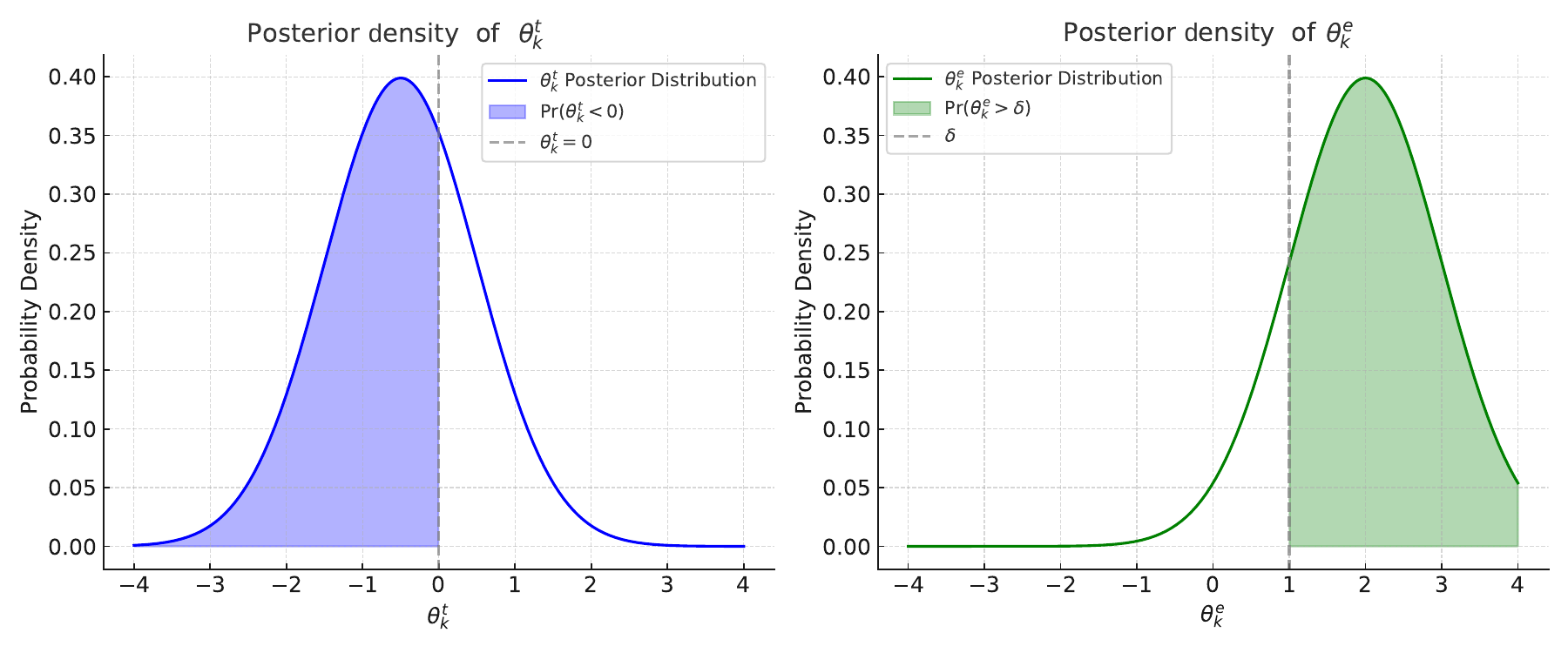}
	\end{center}
	\caption{An example of the separate effect rule. The desirable treatment should have the area of light blue and green shaded parts greater than $\eta_1$ and $\eta_2$ respectively.}
	\label{fig:ind_rule}
\end{figure}

Firstly, a Go decision will only be taken for subtrial $k$ if both (marginal) toxicity and efficacy constraints are satisfied:
$$
\text{Pr}(\theta_{k}^t < 0 \mid \boldsymbol{\mathcal{D}}) > \eta_1\quad\text{and} \quad\text{Pr}(\theta_{k}^e > \delta  \mid \boldsymbol{\mathcal{D}}) > \eta_2,
$$
where $\boldsymbol{\mathcal{D}}$ denotes the toxicity and efficacy data of all subtrials $k=1,\dots, K$; $\eta_1$ and $\eta_2$ are probability thresholds close to 1, and $\delta$ is the expected efficacy difference from the trial specialists. Figure \ref{fig:ind_rule} illustrates the posterior interval probabilities involved in this decision criterion.

This rule aims to make Go decisions only if we have enough belief (posterior probability bigger than a threshold) for toxicity and efficacy separately, which can be regarded as a combination of two marginal rules. However, this decision procedure does not completely conform to our objective of jointly evaluating treatment effects and modelling their possible correlations.

\begin{figure}[ht]
	\begin{center}
		\includegraphics[width=.45\textwidth]{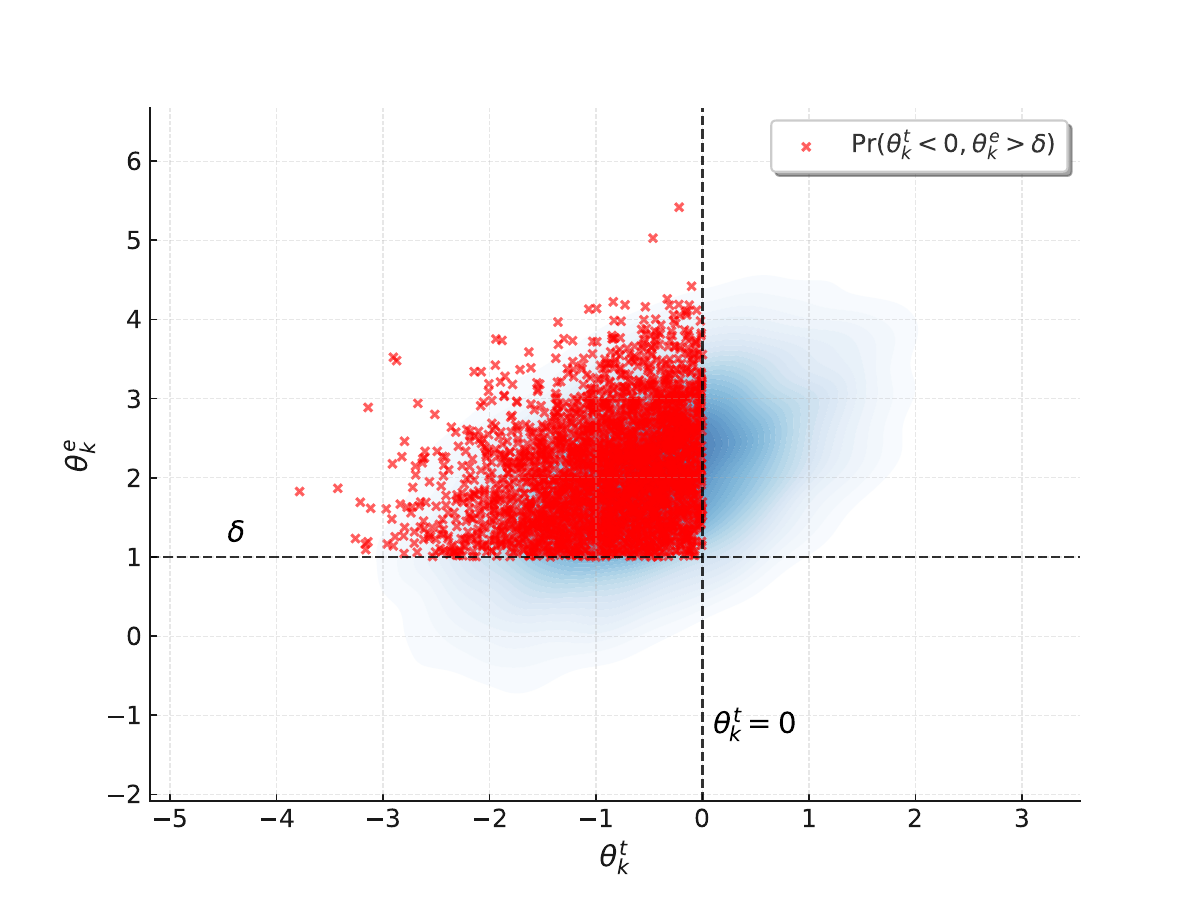}
	\end{center}
	\caption{An example of the joint decision rule. The blue shaded area is the probability density of $(\theta_k^t, \theta_k^e)$. For the desirable treatment, we aim to find the proportion of red crosses in all posterior samples of $(\theta_k^t, \theta_k^e)$ bigger than $\eta$.}
	\label{fig:joint_rule}
\end{figure}

We then propose a joint decision rule involving correlations and only one probability cut-off:
$$
\text{Pr}(\theta_{k}^t < 0 \text{ and } \theta_{k}^e > \delta  \mid \boldsymbol{\mathcal{D}}) > \eta
$$
where $(\theta_k^t, \theta_k^e)$ is assessed in combination to include their correlation. Figure \ref{fig:joint_rule} presents an example of the joint rule in a decision-making procedure.



\section{Simulation Study} \label{sec:sim}

In this section, we evaluate the operating characteristics of basket trials analysed using the proposed Bayesian models with simulations. We simulate basket trial data with $K=6$ subtrials with the data-generation mechanism described in Section \ref{sec:data_generation} and simulation scenarios elaborated in Section \ref{sec:sim_settings}. 

\subsection{Data-generation mechanism} \label{sec:data_generation}

Based on \citet{pm_continuous}, we design a data-generation mechanism to simulate correlated observations of the toxicity and efficacy data from the same individual. This underpins the potential correlation between treatment effects $\theta_k^t$ and $\theta_k^e$ within each subtrial $k$.
This data-generation mechanism is elaborated with three steps below.

\begin{enumerate}
	\item Generate bivariate patient-level data for the control group (C) in subtrial $k$: for patient $i$, generate $(x_{iCk}, z_{iCk})^T \sim \mathcal{N}(\bm{\mu}_{Ck}^\prime, \bm{\Sigma}_{Ck}^\prime)$, where $x_{iCk}$ is a latent variable for generating toxicity outcome, and $z_{iCk}$ is the efficacy observation, $\bm{\mu}_{Ck}^\prime=(\mu_{1Ck}^\prime, \mu_{2Ck}^\prime)^T$ and $ \bm{\Sigma}_{Ck}^\prime$ symbolise the mean and covariance matrix. In this simulation study, we set $\mu_{1C1}^\prime,\cdots,\mu_{1C6}^\prime=(1.2, 1.5, 1.02, 0.88, 1.05, 0.96)$, $\mu_{2Ck}^\prime = \mu_{1Ck}^\prime + 2.0$, and $\bm{\Sigma}_{Ck}^\prime =
	\begin{pmatrix}
		\sigma_{1Ck}^2 & \rho^\prime \sigma_{1Ck}\sigma_{2Ck}\\
		\rho^\prime \sigma_{1Ck}\sigma_{2Ck} & \sigma_{2Ck}^2
	\end{pmatrix} = 
	\begin{pmatrix}
		1^2 & 0.8\times 1\times 1 \\
		0.8\times 1\times 1 & 1^2
	\end{pmatrix}$ for each $k$. 
	\item Dichotomise $x_{iCk}$: Given a fixed threshold $T$, let patient-level toxicity outcomes $y_{iCk} = \text{I}(x_{iCk} \leqslant T)$ such that the binary structure is observable, where I($\cdot$) is an indicator function $= 1$ for toxicity and 0 for no toxicity. It should be noted that $\mu_{2Ck}^\prime$ is the same as $\mu_{Ck}$ in the efficacy observation model described in Section \ref{sec:model_outcomes}. $T$ is set to 0.8 in this simulation study.
	\item Data generation for treatment groups: Suppose we have known $\theta_{k}^t$ and $\theta_{k}^e$ for each subtrial $k$, the mean vector $\bm{\mu}_{Ek}^\prime = (\mu_{1Ek}^\prime, \mu_{2Ek}^\prime)$ can be easily computed via the observation models. We have the identity link for average efficacy responses, hence $\mu_{2Ek}^\prime=\mu_{2Ck}^\prime + \theta_k^e$ ($=\mu_{Ek}$ in Section \ref{sec:model_outcomes}). For obtaining $\mu_{1Ek}^\prime$, we can approximate it by using numerical methods. First, the theoretical toxicity response rate in the treatment group (E) can be computed as
	$$
	p_{Ek} = \text{Pr}(x_{iEk} \leqslant T) = \Phi(T; \mu_{1Ek}^\prime, \sigma_{1Ek}) = 1 - \Phi(\mu_{1Ek}^\prime; T, \sigma_{1Ek})
	$$
	where $x_{iEk}$ is the patient-level latent variable of toxicity in treatment groups, assumed to follow a normal distribution with mean $\mu_{1Ek}^\prime$ and standard deviation $\sigma_{1Ek}$; and $\Phi(x;a,b)$ is the cumulative distribution function (CDF) of $\mathcal{N}(a, b^2)$ at $x$. Therefore,
	\begin{equation} \label{eq:mu1ek}
		\mu_{1Ek}^\prime = \Phi^{-1}(1 - p_{Ek}; T, \sigma_{1Ek})
	\end{equation}
	where $\Phi^{-1}(\cdot)$ is the inverse function of $\Phi(\cdot)$, which explicitly exists and is unique because $\Phi(\cdot)$ is a strictly monotonic function on the real line. Fixed $\sigma_{1Ek}$ at 1, we only need to derive the value of $p_{Ek}$ to attain $\mu_{1Ek}^\prime$. Likewise, the toxicity response rate in the control group
	$$
	p_{Ck} = \text{Pr}(x_{iCk} \leqslant T) = \Phi(T; \mu_{1Ck}^\prime, \sigma_{1Ck})
	$$
	where $\mu_{1Ck}^\prime$ and $\sigma_{1Ck}(=1)$ have been discussed in step 1. Applying the logit link for toxicity in Section \ref{sec:model_outcomes}, we have
	$$
	p_{Ek} = \text{logit}^{-1}(\text{logit}(p_{Ck}) + \theta_k^t)
	$$ Then $\mu_{1Ek}^\prime$ can be obtained via equation \eqref{eq:mu1ek}. Now pseudo-observations for treatment groups can be generated from $(x_{iEk}, z_{iEk})^T \sim \mathcal{N}(\bm{\mu}_{Ek}^\prime, \bm{\Sigma}_{Ek}^\prime)$, with
    \[
    \bm{\mu}_{Ek}^\prime = (\mu_{1Ek}^\prime, \mu_{2Ek}^\prime)^T \quad \text{and} \quad \bm{\Sigma}_{Ek}^\prime = 
	\begin{pmatrix}
		\sigma_{1Ek}^2 & \rho^{\prime\prime} \sigma_{1Ek}\sigma_{2Ek}\\
		\rho^{\prime\prime} \sigma_{1Ek}\sigma_{2Ek} & \sigma_{2Ek}^2
	\end{pmatrix} = 
	\begin{pmatrix}
		1^2 & 0.8\times 1\times 1 \\
		0.8\times 1\times 1 & 1^2
	\end{pmatrix}.
    \] 
    The continuous efficacy and binary toxicity outcomes on the treatment group across subtrials will be $z_{iEk}$ and $y_{iEk} = \text{I}(x_{iEk} \leqslant T)$ respectively.
\end{enumerate}

An important property of the trial outcomes is the induced correlation between the binomial toxicity and continuous efficacy endpoints \citep{cook_1994, Abola2014}. Following our data generation mechanism, the outcome correlation for patients within same subtrial $k$ and treatment group $j$ is
$$
\begin{aligned}
	\text{Cor}(y_{ijk}, z_{ijk}) &= \dfrac{E(y_{ijk} z_{ijk}) - E(y_{ijk})E(z_{ijk})}{\sqrt{\text{Var}(y_{ijk})}\sqrt{\text{Var}(z_{ijk})}} \\
	&= \dfrac{1}{\sigma_{2jk}} \dfrac{\iint \limits_{\mathbb{R}^2} z I(x\leqslant T)\phi\{(x, z)^T;\bm{\mu}_{jk}^\prime, \bm{\Sigma}_{jk}^\prime\}dxdz - \mu_{2jk}^\prime\Phi(T; \mu_{1jk}^\prime, \sigma_{1jk})}{\sqrt{\Phi(T; \mu_{1jk}^\prime, \sigma_{1jk}) - \Phi(T; \mu_{1jk}^\prime, \sigma_{1jk})^2}}
\end{aligned}
$$ where $\phi(\cdot)$ is the probability density function of $\mathcal{N}(\bm{\mu}_{jk}^\prime, \bm{\Sigma}_{jk}^\prime)$. From this equation, the correlation between outcomes can be controlled by adjusting the threshold values $T$, but the detailed adjustment is beyond the scope of this paper and hence will not be discussed.

\subsection{Basic settings} \label{sec:sim_settings}

We simulate the basket trial data assuming equal randomisation between $E$ and $C$ within each subtrial $k = 1, ..., 6$. The subtrial sample sizes are set as: $N_k = \{20, 20, 20, 12, 24, 20\}$ with $n_{Ck} = n_{Ek} = \frac{1}{2}N_k$. We compare our analysis models with their alternatives as follows.

\begin{itemize}
	\item BHM: Standard Bayesian hierarchical model, equivalent to setting each $\omega_{k}=1$ in BiEXNEX.
	\item IndEXNEX: Independent EXNEX models fitted to toxicity and efficacy data. This is a special version of our proposed E-BiEXNEX method without correlations.
	\item SA: Stand-alone analysis, equivalent to setting all $\omega_{k}=0$  and zero-correlation in BiEXNEX for all  $k = 1,\dots, K$.
\end{itemize}

The comparison will be made using Bayesian analogues of the frequentist operating characteristics such as type I error rates and statistical power. This can provide consistent evidence as exploiting the frequentists' approaches, and simplifying complex trial designs to facilitate regulatory compliance \citep{Golchi_Bayesian_metrics}:
\begin{itemize}
	\item (Bayesian) Type I error: for subtrial $k$, the probability that an erroneous Go decision is given under no efficacy or high toxicity.
	\item Power: for subtrial $k$, the probability that a Go decision is recommended with the truly safer and efficacious treatment.
\end{itemize}

We consider the following null and alternative scenarios for evaluating the type I error and power for each modelling approach. 
These scenarios are specified to reflect small to large between-subgroup heterogeneity, as well as low or high correlation between the toxicity and efficacy effects.

\subsubsection{Null scenarios} \label{sec:null_scenarios}

Table \ref{tab:null_scenarios} lists a broad range of scenarios with global and mixed null settings for explorations of type I error rates using the bivariate analysis models.

\begin{table}[ht]
	\centering
	\caption{Null scenarios defined by the subtrial-specific toxicity and efficacy treatment effects. Number in the parentheses is $n_{Ck} = n_{Ek}$, so the total subtrial sample size is twice the number.}
	\label{tab:null_scenarios}
	\resizebox{.95\columnwidth}{!}{
		\begin{tabular}{@{}c|c|cccccc|l@{}}
			\toprule
			\multirow{2}{*}{Scenario} & \multirow{2}{*}{} & \multicolumn{6}{c|}{Subtrial $k$} & \multirow{2}{*}{Description} \\ 
			& & 1 (10) & 2 (10) & 3 (10) & 4 (6) & 5 (12) & 6 (10) & \\ \midrule
			\multirow{2}{*}{Global Null} 
			& $\theta_k^t$ & 0 & 0 & 0 & 0 & 0 & 0 & $\theta_k^t$ and $\theta_k^e$ are both 0 for all subtrials \\ 
			& $\theta_k^e$ & 0 & 0 & 0 & 0 & 0 & 0 & \\ \midrule
			\multirow{2}{*}{1a} 
			& $\theta_k^t$ & 0 & 0 & 0 & 0 & 0 & 0 & $\theta_k^e = 0.75$ across all $k$ for much borrowing \\ 
			& $\theta_k^e$ & 0.75 & 0.75 & 0.75 & 0.75 & 0.75 & 0.75 & \\ \midrule
			\multirow{2}{*}{1b} 
			& $\theta_k^t$ & 0 & 0 & 0 & 0 & 0 & 0 & $\theta_k^e = 2.25$ across all $k$ for much borrowing \\ 
			& $\theta_k^e$ & 2.25 & 2.25 & 2.25 & 2.25 & 2.25 & 2.25 & \\ \midrule
			\multirow{2}{*}{1c} 
			& $\theta_k^t$ & 0 & 0 & 0 & 0 & 0 & 0 & $\theta_k^e$ varies largely \\ 
			& $\theta_k^e$ & 0 & 0.75 & 1.75 & 4 & 8 & 12 & \\ \midrule
			\multirow{2}{*}{2a} 
			& $\theta_k^t$ & -1.25 & -1.25 & -1.25 & -1.25 & -1.25 & -1.25 & $\theta_k^t = -1.25$ across all $k$ for much borrowing \\ 
			& $\theta_k^e$ & 0 & 0 & 0 & 0 & 0 & 0 & \\ \midrule
			\multirow{2}{*}{2b} 
			& $\theta_k^t$ & -2.5 & -2.5 & -2.5 & -2.5 & -2.5 & -2.5 & $\theta_k^t = -2.5$ across all $k$ for much borrowing \\ 
			& $\theta_k^e$ & 0 & 0 & 0 & 0 & 0 & 0 & \\ \midrule
			\multirow{2}{*}{2c} 
			& $\theta_k^t$ & 0 & -0.55 & -1.55 & -3.1 & -6.6 & -9.5 & $\theta_k^t$ varies considerably \\ 
			& $\theta_k^e$ & 0 & 0 & 0 & 0 & 0 & 0 & \\ 
			\bottomrule
		\end{tabular}
	}
\end{table}

Based on the above null settings, we consider the borrowing of information for toxicity or efficacy effects solely or both to investigate type I error rates under different situations, so that the operating characteristics of our model can be thoroughly appraised.

\subsubsection{Scenarios for power calculation} \label{sec:power_scenarios}
Six simulation scenarios are presented in Table \ref{tab:power_scenarios} for power calculation. As a succinct summary, the total six scenarios can be roughly divided into three groups:

\begin{itemize}\label{list:theta}
	\item Treatment effects $\theta_{k}^t$ and $\theta_{k}^e$ are similar or partly similar: scenario Ia, Ib
	\item $\theta_{k}^t$ and $\theta_{k}^e$ are largely different: scenario IIa (correlation as low as -0.18 between $\theta_1^t, \theta_2^t,\cdots, \theta_K^t$ and $\theta_1^e, \theta_2^e, \cdots, \theta_K^e$) and IIb (high negative correlation as -0.94)
	\item One of $\theta_k^t$ and $\theta_{k}^e$ is similar while the other is not: scenario IIIa and IIIb
\end{itemize}

\noindent where the similarity measure of treatment effects is their standard deviation (SD), that is, SD $>$ 1 means the effects are considered largely different. We include the accordingly transformed toxicity response rates and mean efficacy for treatment and control groups in supplementary material \ref{app:res_rates} for reference.

\begin{table}[ht]
	\centering
	\caption{Six additional scenarios for power calculation. The first row indicates toxicity treatment effects, the second row shows efficacy treatment effects, and the last column provides simple descriptions. The middle column SD gives the standard deviation of treatment effects accordingly.}
	\label{tab:power_scenarios}
	\resizebox{\textwidth}{!}{
		\scriptsize
		\begin{tabular}{@{}c|c|cccccc|c|l@{}}
			\toprule
			\multirow{2}{*}{Scenario} & \multirow{2}{*}{} & \multicolumn{6}{c|}{subtrial $k$ with (half-)sample size $n_{Ck}=n_{Ek}$} & \multirow{2}{*}{SD} & \multirow{2}{*}{Description} \\ 
			& & 1 (10) & 2 (10) & 3 (10) & 4 (6) & 5 (12) & 6 (10) &  & \\ \midrule
			\multirow{2}{*}{Ia} 
			& $\theta_k^t$ & -0.8  & -0.68 & -0.72 & -0.77 & -0.96 & -0.92 & 0.11 & $\theta_k^t$ and $\theta_k^e$ are similar across subtrials \\ 
			& $\theta_k^e$ & 0.76  & 0.82  & 0.69  & 0.72  & 0.75  & 0.66  & 0.06 & \\ \midrule
			\multirow{2}{*}{Ib} 
			& $\theta_k^t$ & -0.95 & -0.82 & -1.07 & 0.20  & -1.67 & -2.10 & 0.79 & $\theta_k^t$ and $\theta_k^e$ are partly similar across subtrials \\ 
			& $\theta_k^e$ & 0.77  & 0.73  & 0.80  & 1.80  & -0.46 & -1.20 & 1.06 & \\ \midrule
			\multirow{2}{*}{IIa} 
			& $\theta_k^t$ & -1.4  & 1.4   & 0.87  & -2.2  & 0.15  & -0.8  & 1.38 & $\theta_k^t$ and $\theta_k^e$ are largely different, with low correlation (-0.18) \\ 
			& $\theta_k^e$ & -1.77 & 0.05  & 0.68  & 1.80  & -0.46 & 2.12  & 1.45 & \\ \midrule
			\multirow{2}{*}{IIb} 
			& $\theta_k^t$ & -1.4  & 1.4   & -0.87 & -2.2  & 0.15  & 0.8   & 1.37 & $\theta_k^t$ and $\theta_k^e$ are largely different, with high correlation (-0.94) \\ 
			& $\theta_k^e$ & 1.43  & -1.66 & 1.08  & 1.46  & -0.02 & -1.84 & 1.51 & \\ \midrule
			\multirow{2}{*}{IIIa} 
			& $\theta_k^t$ & -1.5  & 2.34  & 0.87  & -0.55 & -3    & 4.22  & 2.63 & $\theta_k^t$ is similar, but $\theta_k^e$ is not \\ 
			& $\theta_k^e$ & 1.0   & 1.0   & 1.0   & 1.0   & 1.0   & 1.0   & 0    & \\ \midrule
			\multirow{2}{*}{IIIb} 
			& $\theta_k^t$ & -1.2  & -1.2  & -1.2  & -1.2  & -1.2  & -1.2  & 0    & $\theta_k^e$ is similar, but $\theta_k^t$ is not \\ 
			& $\theta_k^e$ & -1.27 & 3.48  & 2.78  & -0.02 & 1.15  & -0.12 & 1.83 & \\ 
			\bottomrule
		\end{tabular}
	}
\end{table}

\subsection{Prior specification} \label{sec:sim_prior}

We assume no specific opinion on exchangeability and non-exchangeability for toxicity and efficacy treatment effects, so all related prior probabilities of exchangeability would be set to $\omega_k=0.5$ in BiEXNEX, and $\lambda_{ik}=0.25$ in E-BiEXNEX for each of the four component and subtrial $k$. In order to make models comparable, the identical priors of corresponding parameters for each simulated method will be employed. For the observation model, we assume $\alpha_k^t$ and $\alpha_{k}^e$ follow a uninformative prior $\mathcal{N}(0, 10^2)$ for all subtrials, which can accommodate a large range of possible baseline effects, together with a vague variance prior of efficacy observations $\sigma_{jk}^2 \sim \text{Half-}\mathcal{N}(5^2)$, where $\text{Half-}\mathcal{N}(x)$ means the half-normal distribution with variance $x$ folded at mean 0.

Vague priors are placed on $\beta_1$ and $\beta_{2}$: $\beta_1, \beta_2 \sim \mathcal{N}(0, 5^2)$. We set $\phi_1^2, \phi_2^2\sim \text{Half-}\mathcal{N}(0.5^2)$ because of the need to constrain the upper value of such variances when clinically relevant differences are assumed to be exchangeable, and the scale parameter 0.5 was chosen to allow medium to substantial between-subtrial heterogeneity \citep{exnex}. On the other hand, subtrial specific parameters $m_{1k}, m_{2k}=0$ and $s_{1k}, s_{2k}=5$ for all subtrials, and the unknown correlations $\rho, \kappa$ are presupposed to follow a flat prior $\mathcal{U}(-1,1)$.

We use Markov Chain Monte Carlo (MCMC) method to simulate each model (not generate the data) and record the posterior samples of treatment effects, in which the median is adopted as the point estimate as it is robust to skewed posterior densities. To control the Monte Carlo standard error under 0.5\%, we perform $10^4$ repetitions in the study \citep{morris} for each model and scenario on a computer with R4.3.2 and Just Another Gibbs Sampler (JAGS4.3.1). We have four MCMC chains for every model in a single repetition, each with 2000 burn-in samples and $10^4$ iterations.

\subsection{Results} \label{sec:sim_results}

All figures and main results in this section are presented by applying the joint decision rule (Section \ref{sec:dmrule}), because of negligible difference between the simulated results by both rules. 

\begin{figure}[ht]
	\begin{center}
		\includegraphics[width=\textwidth]{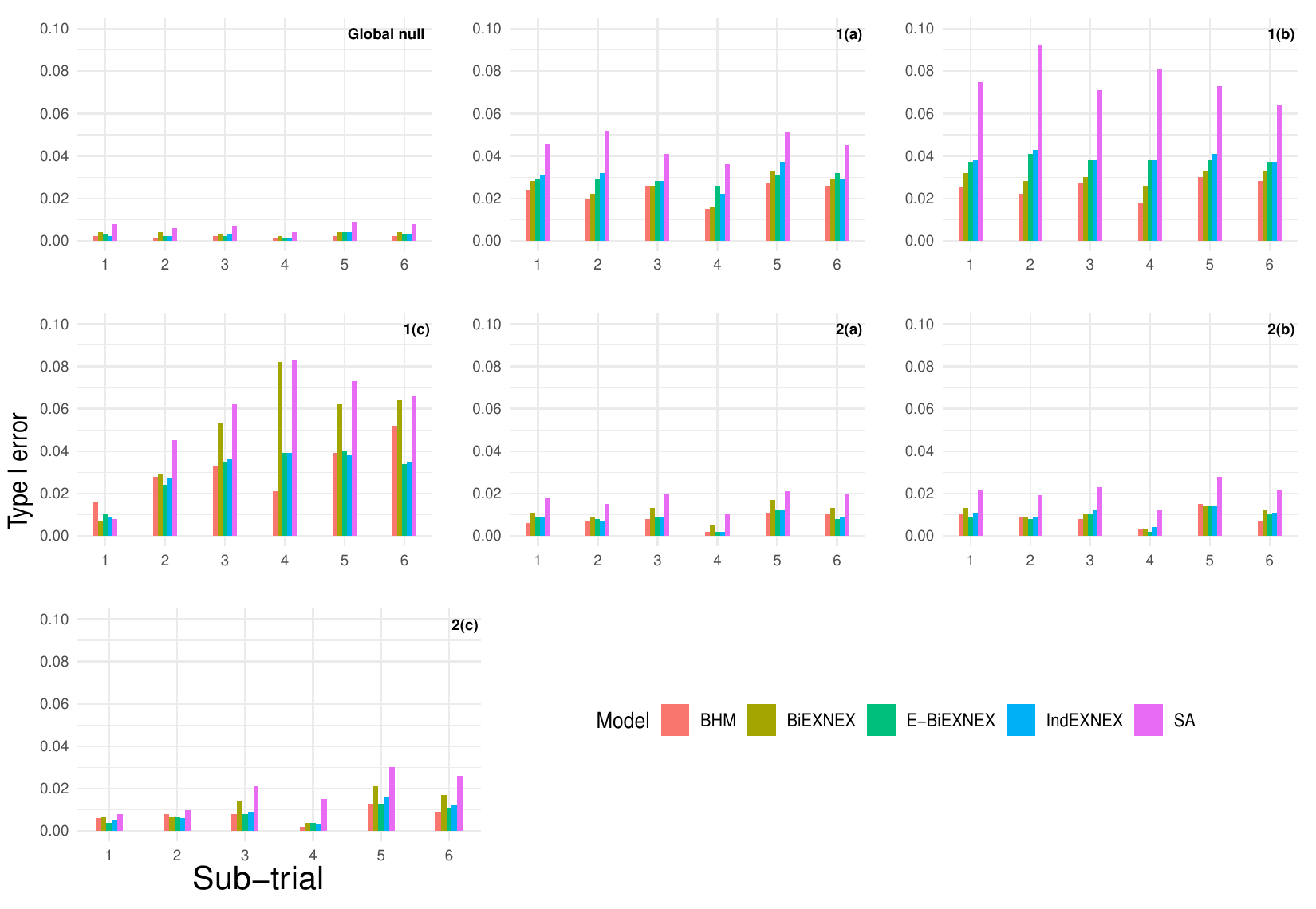}
	\end{center}
	\caption{Bayesian analogues of type I error rates for each null scenario in Section \ref{sec:null_scenarios}. The x-axis is subtrial index and y-axis is type I error rate for each model. 
    }
	\label{fig:typeIerr}
\end{figure}

Figures~\ref{fig:typeIerr} shows the type I error rates for all models under various null scenarios. 
Under the global null scenario (top left panel of Figure~\ref{fig:typeIerr}), all methods, including BHM, BiEXNEX, E-BiEXNEX, IndEXNEX, and SA, demonstrate small and comparable type I error rates. This indicates that, where there is no toxicity and efficacy difference between the treatment and control group, all approaches perform similarly well with low error rates.

In the mixed null scenarios 1a -- 1c, where $\theta_k^t = 0$ but $\theta_k^e$ increases (from 1a to 1b) or varies (1c), noticeable differences emerge. The stand-alone analysis (SA) method consistently produces the highest type I error rates compared with approaches borrowing information, because it generates wider posterior distributions of treatment effects without borrowing information across patient subgroups, causing higher type I error rates than other methods exploiting more information. Although BHM and BiEXNEX model performs adequately well when both toxicity and efficacy treatment effects are exchangeable, their type I error rates can fluctuate to a significant extent when only one effect is exchangeable. The two models perform well in scenarios 1a and 1b due to their capability to correctly specify the substantial degree of information borrowing due to identical $\theta_k^t$ and $\theta_k^e$ across all sub-trials. While BHM and BiEXNEX produce largely varied type I error in scenario 1c, where only $\theta_k^t$ is exchangeable, since they cannot accommodate the different exechangeability of treatment effects. To be exact, in scenario 1c, the BHM method performs well in subtrials 1 -- 4 but generate rapidly increasing error in subtrials 5 and 6 as it borrows too much information from subtrials 1 -- 4 which have more similar $\theta_k^e$ in contrast with subtrials 5 and 6. BiEXNEX model produces larger error rates as it cannot specify the correct pattern of information borrowing, but the type I error of BiEXNEX is still less than SA (it is natural for BiEXNEX as a combination of SA and BHM).

In scenarios 2a--2c, the SA method again exhibits the largest type I error rates. The other models, such as BiEXNEX, E-BiEXNEX, and IndEXNEX, maintain small and stable error rates even in scenario 2c, where $\theta_k^t$ values are highly varied. The reason for this phenomenon is that, on the logit scale, $\theta_k^t = -3.1, -6.6, -9.5$ actually impact less on the simulated data (e.g. if $p_{Ck} = 0.3$, $p_{Ek} = \text{logit}^{-1}(\text{logit}(p_{Ck}) + \theta_k^t) \approx 0.02, 6\times10^{-4}, 3\times10^{-5}$ which will have very similar effects in data generation under the current limited sample size). Therefore, there would be more information shared in scenario 2c compared with 1c, to produce similar type I error rates for methods excluding SA. Besides, the probability of making erroneous decisions for at least one subtrial (defined as overall error rate, OER) is shown in supplementary material \ref{app:oer} as a referential statistics if the "family-wise error rates" are considered.

\begin{figure}[ht]
	\begin{center}
		\includegraphics[width=\textwidth]{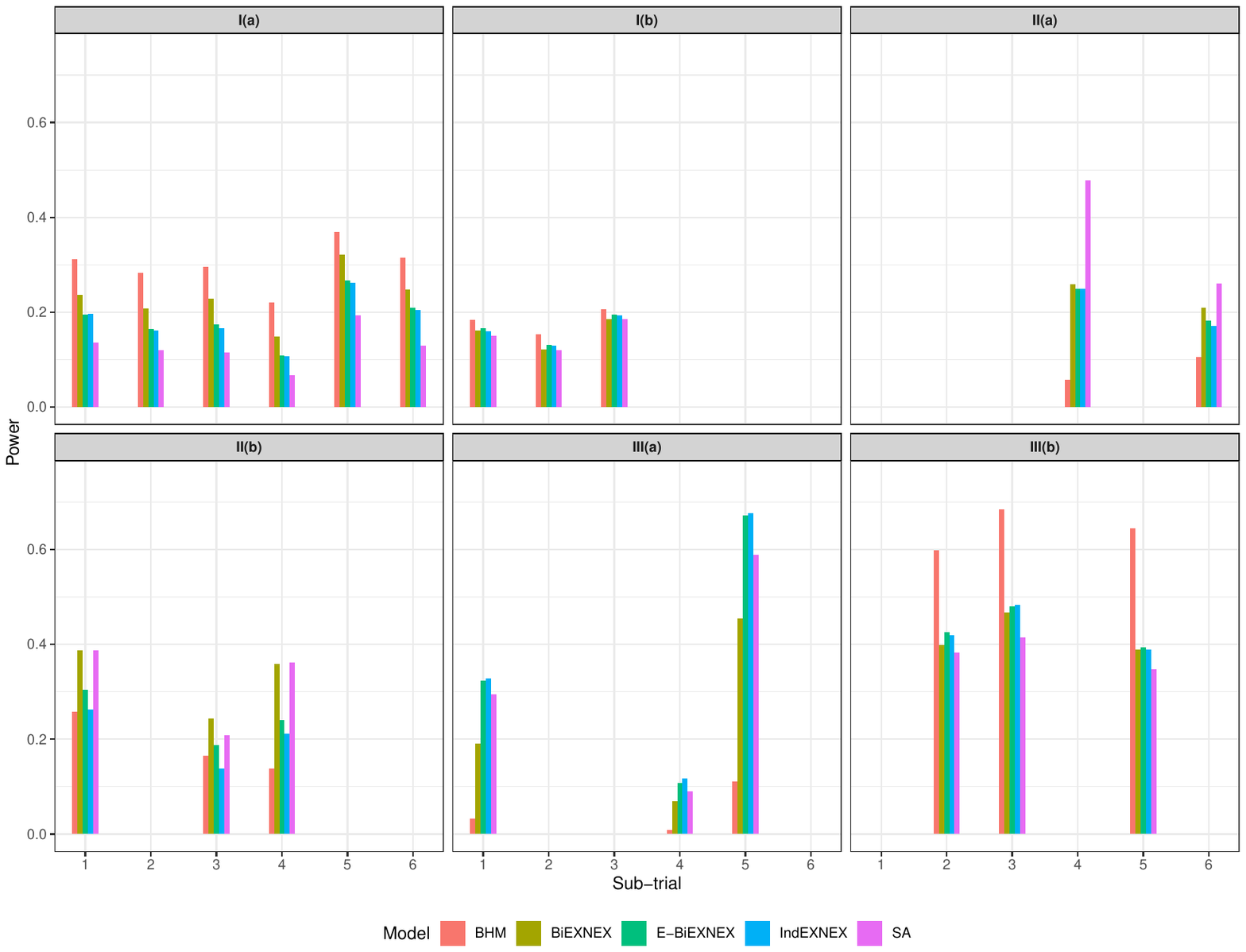}
	\end{center}
	\caption{Bayesian analogues of power for each scenario, model, and subtrial. The Go decision proportions under the subtrials with overly toxicity or inefficacy are erased from the figure. 
    }
	\label{fig:power}
\end{figure}

Figure~\ref{fig:power} illustrates the statistical power for all models under each scenario. With differing levels of similarities between treatment effects, all approaches have different efficiency and robustness. When the dissimilarity of treatment effects grows (e.g. from scenario Ia to IIa), the BHM performs more inefficient while SA produces larger power. In scenario Ia, where treatment effects are similar across subtrials, the BHM method achieves the highest power. This performance is expected due to the effective information sharing enabled by the small variability in treatment effects. However, in scenarios IIa and IIb, the performance of BHM is unsatisfactory because of the excessive borrowing of information from considerably varied treatment effects. 
By contrast, SA which dose not assume any information borrowing, performs inverse but analogously extreme compared to BHM. Other methods are able to produce robust power in scenarios Ia, Ib, IIa and IIb as they consider the possibility of the high level of information borrowing or non-exchangeable treatment effects.

The obvious correlation between treatment effects also have noticeable influence on model performances, especially when both treatment effects vary substantially (scenarios IIa and IIb). It is clear that in scenario IIb where treatment effects are largely different with high correlation, BiEXNEX can perfectly specify the scenario, leading to very similar power in subtrial 1 and 4, and even higher power in subtrial 3, compared with SA which does not consider possible correlation. 
On average, E-BiEXNEX produces about 5\% higher power than IndEXNEX in scenario IIb as it accommodates such a correlation.

In scenarios IIIa and IIIb where only $\theta_k^t$ or $\theta_k^e$ varies dramatically, BHM generates lower power than SA in IIIa but higher in IIIb under the current sample sizes, while E-BiEXNEX and IndEXNEX models produce similar power. In scenario IIIa, where $\theta_k^e$ is the same but $\theta_k^t$ changes significantly, E-BiEXNEX and IndEXNEX achieve substantially high power, reflecting their ability to account for the exchangeability of sole effect. In comparison, BHM produces the lowest power, falling below that of SA. In scenario IIIb, where $\theta_k^t$ is identical and $\theta_k^e$ varies, BHM performs exceptionally well, achieving the highest power, while SA reports the lowest power, i.e., below 0.4 in most cases. The reason lies in the current sample sizes, which can be sufficient to infer efficacy-related parameters from continuous data, but far not enough to obtain precise toxicity treatment effect for binary toxicity data.
This aligns well with the statement in \citet{Altman2006} that less information is carried in dichotomised data compared to continuous data. Therefore, in scenario IIIb, the same $\theta_k^t$ indicates borrowing of information could be recommended under the current sample size. 

We also investigate the scenarios with only one exchangeable treatment effects with smaller sample sizes (half of the current sizes). The results show E-BiEXNEX achieves the highest power, regardless which treatment effects, on either toxicity or efficacy, are exchangeable, when the smaller samples cannot provide sufficient information even in continuous data. 
Relevant figures are included in supplementary material \ref{app:smaller_samples}.
For results obtained by applying the first decision criterion concerning marginal effects, we refer the interested readers to  supplementary material \ref{app:sim_results}. 

Additionally, we investigate the impact of correlations between treatment effects on the posterior weights of our models that permit information borrowing. It was found that when $\theta_k^t$ and $\theta_k^e$ are highly correlated (e.g., the correlation is near -0.9) and largely different (so less information borrowing from their differences), the E-BiEXNEX model will give higher (3 -- 8 \%) posterior weight to ``both treatment effects are non-exchangeable" compared with using IndEXNEX. However, when $\theta_{k}^t$ and $\theta_k^e$ both are very similar, the correlation will have nearly no impact on the posterior weights. Therefore, the similarity between treatment effects is the main factor to influence the information borrowing weights, while the treatment effect correlation does have some impact on the posterior weight under low similarity and highly correlated effects scenarios. 

The results highlight clear trade-offs between the models and the appealing robustness of E-BiEXNEX which demonstrates steady performance across all scenarios, while achieving low type I error rates and satisfactory statistical power. SA, while providing larger type I error rates, performs acceptably in scenarios with substantial heterogeneity in treatment effects given the current large sample sizes. BHM excels when treatment effects are highly similar but struggles in scenarios with large variability in the toxicity and efficacy effects across subtrials. IndEXNEX achieve balanced performance, particularly in cases involving only one exchangeable effect, while it loses to E-BiEXNEX when varied treatment effects are highly correlated. The BiEXNEX method strikes a balance between BHM and SA, but it can also miss situations with solely exchangeable treatment effects. Overall, E-BiEXNEX is shown to be the most robust method, delivering consistently small type I error rates and competitive power across all tested scenarios.


\section{Discussion} \label{sec:discuss}

Continued toxicity monitoring in phase II clinical trials gains increasing attention in drug development to recommend not only efficacious but safe treatments for further investigation. 
Basket trials spark natural possibilities of information borrowing across subtrials, establishing a joint analysis of efficacy and toxicity while accounting for patient heterogeneity in settings of precision medicine. 
In this paper, we have proposed two bivariate Bayesian analysis models to this end, with a comprehensive simulation study conducted to evaluate the trial operating characteristics for each model and their alternatives.

From the simulation study in Section \ref{sec:sim}, 
BiEXNEX and E-BiEXNEX are more robust than BHM and SA across all scenarios, and E-BiEXNEX can flexibly accommodate scenarios in which only a single treatment effect is exchangeable, thus producing greater power. At the same time, small and stable type I error rates are achieved by E-BiEXNEX in all null scenarios. Correlation between treatment effects can have some impact on power calculation, but the similarity is the main factor. We conclude that when both treatment effects are assumed to be similar or largely different, BiEXNEX model is recommended to achieve high power and safe type I error rates; when asynchronous exchangeability or no prior knowledge is presupposed, E-BiEXNEX is a more appropriate choice to realise robust decisions.

One extension of our proposed methodology lies in the number of exchangeability distributions that can adequately accommodate the vectors of toxicity and efficacy parameters, 
and how assumptions of exchangeability can be relaxed for cases where some subtrials have more similar treatment effects between themselves than with others.
Inspirations may be drawn from \citet{hz_discrepancy} that enables pairwise borrowing of information and \citet{JackLee_BCHM} which involves clustering of subgroups for efficient information borrowing. 
Another research avenue to pursue is to develop sample size determination that accounts for the joint evaluation of efficacy and toxicity, with borrowing of information across subtrials \citep{zheng2023}. 
Finally, we are interested in exploring the joint evaluation of toxicity and efficacy in basket trials where simultaneous dose-finding proceduress may be implemented across patient subgroups. 
One possibility is to extend our Bayesian models to facilitate the estimation of dose-toxicity and dose-response curves in such settings, with the principal aim for identification of the optimal biological dose \citep{Mandrekar2010_OBD}. Another direction in dose-finding settings is the consideration for late-onset toxicity which is common in clinical practice \citep{yz_late_onset_review}.

\section*{Acknowledgement}

ZC receives the doctoral training scholarship for PhD in Biostatistics by the MRC Trials Methodology Research Partnership Doctoral Training Partnership (Grant Ref: MR/W006049/1). This report is an independent research supported by the National Institute for Health Research (NIHR300576). The views expressed in this publication are those of the authors and not necessarily those of the NHS, the National Institute for Health
Research or the Department of Health and Social Care (DHSC). PM also received funding from UK Medical Research Council (MC UU 00002/19). 
Dr Zheng's contribution to this manuscript was supported by Cancer Research UK (RCCCDF-May24/100001). 
For the purpose of open access, the author has applied a Creative Commons Attribution (CC BY) licence to
any Author Accepted Manuscript version arising. 

\appendix

\newpage
\section*{Supplementary Material}
\renewcommand\thefigure{\thesection\arabic{figure}}    

\section{Another model stipulation for E-BiEXNEX} \label{app:another_stipulation}
\label{appendix:sepE-BiEXNEX}

For subtrial $k$,
\begin{itemize}
	\item $\mathcal{C}(\cdot)$: categorical distribution
	\item $\mathcal{N}(\cdot)$: two-dimensional normal distribution
	\item $\text{Z}_{tk}$ or $\text{Z}_{ek}$: the exchangeability indicators as auxiliary variables, where t means toxicity and e denotes efficacy
\end{itemize}
\begin{equation}\label{eb1}
	\begin{aligned}
		\text{Z}_{tk} &\sim \mathcal{C}(\bm{\pi}_{tk}) \\
		\text{Z}_{ek} &\sim \mathcal{C}(\bm{\pi}_{ek}) \\
		\left.
		\begin{pmatrix}
			\theta_{k}^t \\
			\theta_{k}^e
		\end{pmatrix} \right| \text{Z}_{tk} = i, \text{Z}_{ek} &= j \sim \mathcal{N}(\bm{\mu}_{ijk}, \bm{\Sigma}_{ijk})
	\end{aligned}
\end{equation}

where i and j can be 1 (exchangeable) or 2 (non-exchangeable); $\mathcal{N}(\cdot,\cdot)$ is the two-dimensional distribution with mean $\bm{\mu}_{ijk}$ and covariance matrix $\bm{\Sigma}_{ijk}$ which can represent the parameters of each component in the E-BiEXNEX (Section \ref{sec:ebiexnex}) model (i.e.
$$
(\bm{\mu}_{11k}, \bm{\Sigma}_{11k}) = 
\left[
\begin{pmatrix}
	\beta_1 \\
	\beta_2
\end{pmatrix},
\begin{pmatrix}
	\phi_1^2 & \rho\phi_1\phi_2 \\
	\rho\phi_1\phi_2 & \phi_2^2
\end{pmatrix}\right];\qquad
(\bm{\mu}_{12k}, \bm{\Sigma}_{12k}) = 
\left[\begin{pmatrix}
	\beta_1 \\
	m_{2k}
\end{pmatrix},
\begin{pmatrix}
	\phi_1^2 & 0 \\
	0 & s_{2k}^2
\end{pmatrix}\right]
$$
$$
(\bm{\mu}_{21k}, \bm{\Sigma}_{21k}) = 
\left[\begin{pmatrix}
	m_{1k} \\
	\beta_{2}
\end{pmatrix},
\begin{pmatrix}
	s_{1k}^2 & 0 \\
	0 & \phi_2^2
\end{pmatrix}\right]
;\qquad
(\bm{\mu}_{22k}, \bm{\Sigma}_{22k}) =
\left[\begin{pmatrix}
	m_{1k} \\
	m_{2k}
\end{pmatrix},
\begin{pmatrix}
	s_{1k}^2 & \kappa s_{1k} s_{2k} \\
	\kappa s_{1k} s_{2k} & s_{2k}^2
\end{pmatrix}\right]
$$
) $\bm{\pi}_t=(\omega_t, 1-\omega_t)$ and $\bm{\pi}_e=(\omega_e, 1-\omega_e)$ with prior exchangeability weights $\omega_t$, $\omega_e$ on toxicity and efficacy respectively.

Now proof of the equivalence between this stipulation and the original one is as follows: First of all, we can similarly write the original E-BiEXNEX model for subtrial $k$ as

\begin{equation}\label{eb2}
	\begin{aligned}
		\text{Z} &\sim \mathcal{C}(\bm{\pi}_k) \\
		\left.
		\begin{pmatrix}
			\theta_{k}^t \\
			\theta_{k}^e
		\end{pmatrix} \right| \text{Z}_k = &l \sim \mathcal{N}(\bm{\mu}_{lk}, \bm{\Sigma}_{lk})
	\end{aligned}
\end{equation}
where $l = 1,2,3,4$ denoting 4 components section \ref{sec:ebiexnex}; $\bm{\mu}_{lk}$ and $\bm{\Sigma}_{lk}$ are corresponding mean and covariance matrices. Apparently $(\bm{\mu}_{1k}, \bm{\Sigma}_{1k})=(\bm{\mu}_{11k}, \bm{\Sigma}_{11k})$, $(\bm{\mu}_{2k}, \bm{\Sigma}_{2k})=(\bm{\mu}_{12k}, \bm{\Sigma}_{12k})$, $(\bm{\mu}_{3k}, \bm{\Sigma}_{3k})=(\bm{\mu}_{21k}, \bm{\Sigma}_{21k})$, $(\bm{\mu}_{4k}, \bm{\Sigma}_{4k})=(\bm{\mu}_{22k}, \bm{\Sigma}_{22k})$, and $\bm{\pi}_k=(\lambda_{1k}, \lambda_{2k}, \lambda_{3k}, \lambda_{4k})$.

In model \ref{eb2}, the posterior exchangeability weight for toxicity can be approximated by Pr$\left(\text{Z}_k = 1 \left| 
\begin{pmatrix}
	\theta_{k}^t \\
	\theta_{k}^e
\end{pmatrix}\right.\right)+\text{Pr}\left(\text{Z}_k = 2 \left| 
\begin{pmatrix}
	\theta_{k}^t \\
	\theta_{k}^e
\end{pmatrix}\right.\right)$ while in model \ref{eb1} it is exactly Pr$\left(\text{Z}_{tk} = 1 \left| 
\begin{pmatrix}
	\theta_{k}^t \\
	\theta_{k}^e
\end{pmatrix}\right.\right)$. Next we show if $\text{Pr}(Z_{k} = 1) = \text{Pr}(Z_{tk} = 1, Z_{ek} = 1)$,$\text{Pr}(Z_{k} = 2) = \text{Pr}(Z_{tk} = 1, Z_{ek} = 2)$, $\text{Pr}(Z_{k} = 3) = \text{Pr}(Z_{tk} = 2, Z_{ek} = 1)$ and $\text{Pr}(Z_{k} = 4) = \text{Pr}(Z_{tk} = 2, Z_{ek} = 2)$ then
\begin{equation}\label{eq:approx}
	\text{Pr}\left(\text{Z}_{tk} = 1 \left| 
	\begin{pmatrix}
		\theta_{k}^t \\
		\theta_{k}^e
	\end{pmatrix}\right.\right)
	=
	\text{Pr}\left(\text{Z}_k = 1 \left| 
	\begin{pmatrix}
		\theta_{k}^t \\
		\theta_{k}^e
	\end{pmatrix}\right.\right)+\text{Pr}\left(\text{Z}_k = 2 \left| 
	\begin{pmatrix}
		\theta_{k}^t \\
		\theta_{k}^e
	\end{pmatrix}\right.\right)
\end{equation}

\textit{Proof.}

According to model \ref{eb1},
$$
\begin{aligned}
	\text{Pr}\left(\text{Z}_{tk} = 1 \left| 
	\begin{pmatrix}
		\theta_{k}^t \\
		\theta_{k}^e
	\end{pmatrix}\right.\right) &= 
	\sum_{r=1}^2\text{Pr}\left(\text{Z}_{tk} = 1, \text{Z}_{ek} = r \left| 
	\begin{pmatrix}
		\theta_{k}^t \\
		\theta_{k}^e
	\end{pmatrix}\right.\right)\\
	&=
	\dfrac{\sum_{r=1}^2\text{Pr}(\text{Z}_{tk} = 1, \text{Z}_{ek} = r)\text{Pr}\left(
		\left.\begin{pmatrix}
			\theta_{k}^t \\
			\theta_{k}^e
		\end{pmatrix}  \right| \text{Z}_{tk} = 1, \text{Z}_{ek} = r\right)}{\sum_{i,j=1}^2\text{Pr}(\text{Z}_{tk} = i, \text{Z}_{ek} = j)\text{Pr}\left(
		\left.\begin{pmatrix}
			\theta_{k}^t \\
			\theta_{k}^e
		\end{pmatrix}  \right| \text{Z}_{tk} = i, \text{Z}_{ek} = j\right)}
\end{aligned}
$$
while model \ref{eb2} shows
$$
\begin{aligned}
	\sum_{r=1}^2\text{Pr}\left(
	\text{Z}_{k} = r  \left|\begin{pmatrix}
		\theta_{k}^t \\
		\theta_{k}^e
	\end{pmatrix}\right.\right)
\end{aligned} = 
\dfrac{\sum_{r=1}^2\text{Pr}(\text{Z}_{k} = r)\text{Pr}\left(
	\left.\begin{pmatrix}
		\theta_{k}^t \\
		\theta_{k}^e
	\end{pmatrix}  \right| \text{Z}_{k} = r\right)}{\sum_{l=1}^4\text{Pr}(\text{Z}_{k} = l)\text{Pr}\left(
	\left.\begin{pmatrix}
		\theta_{k}^t \\
		\theta_{k}^e
	\end{pmatrix}  \right| \text{Z}_{k} = l\right)}
$$
Due to the relation of likelihoods between the two models, it is obvious that if the previous equations for prior probabilities of $Z_{k}$ and $(Z_{tk}, Z_{ek})$ hold, the two sides in equation \ref{eq:approx} will be strictly equal, indicating that the two models are identical under very simple constraints concerning prior exchangeability weights.
\clearpage

\section{The response rates for control and treatment groups under each scenario for power calculation} \label{app:res_rates}

\begin{table}[htbp]
	\centering
	\caption{$p_{Ek}$ specification table with 6 scenarios and $K=6$ subtrials. The bottom row indicates $p_{Ck}$ for all subtrials.}
	\label{tab:sim_tox_rates}
	\resizebox{0.6\columnwidth}{!}{
		\begin{tabular}{@{}c|cccccc|c@{}}
			\toprule
			\multirow{2}{*}{Scenario} & \multicolumn{6}{c|}{subtrial $k$ with (half-)sample size $n_{Ck}=n_{Ek}$, e.g. 1 (10)}       & \multirow{2}{*}{} \\
			& 1 (10)        & 2 (10)        & 3 (10)        & 4 (6)        & 5 (12)        & 6 (10)                                         \\ \midrule
			Ia                         & 0.19  & 0.14 & 0.26 & 0.29 & 0.20 & 0.24                                    \\
			Ib                         & 0.17  & 0.12 & 0.19 & 0.52 & 0.11 & 0.09                                   \\
			IIa                         & 0.11  & 0.56 & 0.63 & 0.09 & 0.43 & 0.26                                   \\
			IIb                         & 0.11  & 0.56 & 0.23 & 0.09 & 0.44 & 0.63                                  \\
			IIIa                         & 0.10  & 0.77 & 0.63 & 0.34 & 0.03 & 0.98                                    \\
			IIIb                         & 0.14  & 0.09 & 0.17 & 0.21 & 0.17 & 0.19                                      \\ \midrule
			$p_{Ck}$                  & 0.34          & 0.24          & 0.41          & 0.46          & 0.40          & 0.43                                         \\ \bottomrule
		\end{tabular}
	}
\end{table}

\begin{table}[htbp]
	\centering
	\caption{$\mu_{Ek}$ specification table with 6 scenarios and $K=6$ subtrials, with the bottom row showing $\mu_{Ck}$ for all subtrials.}
	\label{tab:sim_eff_mean}
	\resizebox{.6\columnwidth}{!}{
		\begin{tabular}{@{}c|cccccc|c@{}}
			\toprule
			\multirow{2}{*}{Scenario} & \multicolumn{6}{c|}{subtrial $k$ with (half-)sample size $n_{Ck}=n_{Ek}$, e.g. 1 (10)}      & \multirow{2}{*}{} \\
			& 1 (10)        & 2 (10)        & 3 (10)       & 4 (6)        & 5 (12)        & 6 (10)        &                                         \\ \midrule
			Ia                         & 3.96 & 4.32 & 3.71 & 3.60 & 3.80 & 3.62 &                                     \\
			Ib                         & 3.97 & 4.23 & 3.82 & 4.68 & 2.59 & 1.76 &                                     \\
			IIa                         & 1.43 & 3.55 & 3.70 & 4.68 & 2.59 & 5.08 &                                     \\
			IIb                         & 4.63 & 1.84 & 4.10 & 4.34 & 3.03 & 1.12 &                                     \\
			IIIa                         & 4.20 & 4.50 & 4.02 & 3.88 & 4.05 & 3.96 &                                        \\
			IIIb                         & 1.93 & 6.98 & 5.80 & 2.86 & 4.20 & 2.84 &                                    \\ \midrule
			$\mu_{Ck}$                & 3.20          & 3.50          & 3.02         & 2.88          & 3.05          & 2.96          &                                         \\ \bottomrule
		\end{tabular}
	}
\end{table}

\clearpage
\section{Overall error rates} \label{app:oer}

\begin{figure}[ht]
	\begin{center}
		\includegraphics[width=0.9\textwidth]{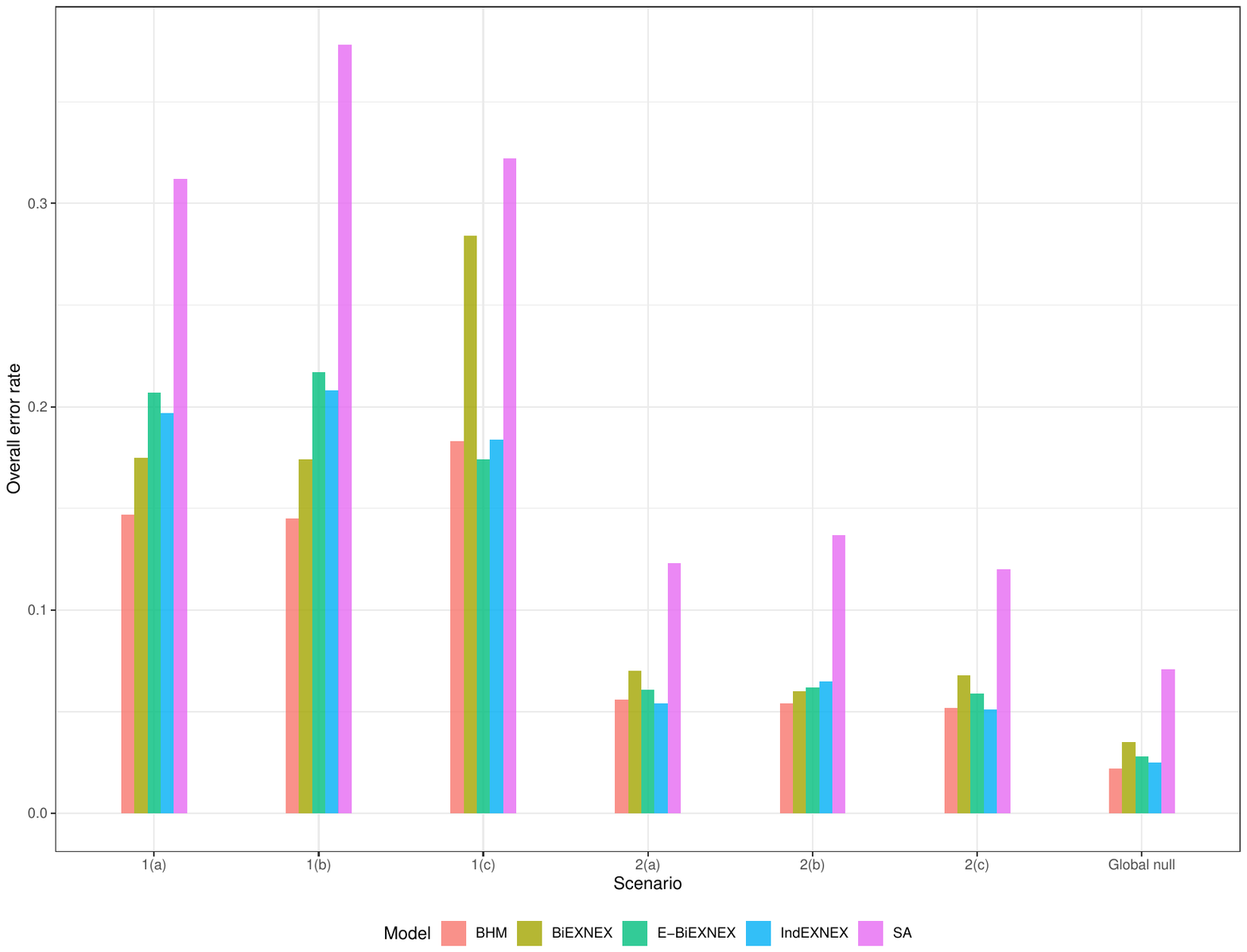}
	\end{center}
	\caption{The overall error rate under each null scenario for the first decision rule.}
	\label{fig:oer_appendix}
\end{figure}

\begin{figure}[ht]
	\begin{center}
		\includegraphics[width=0.9\textwidth]{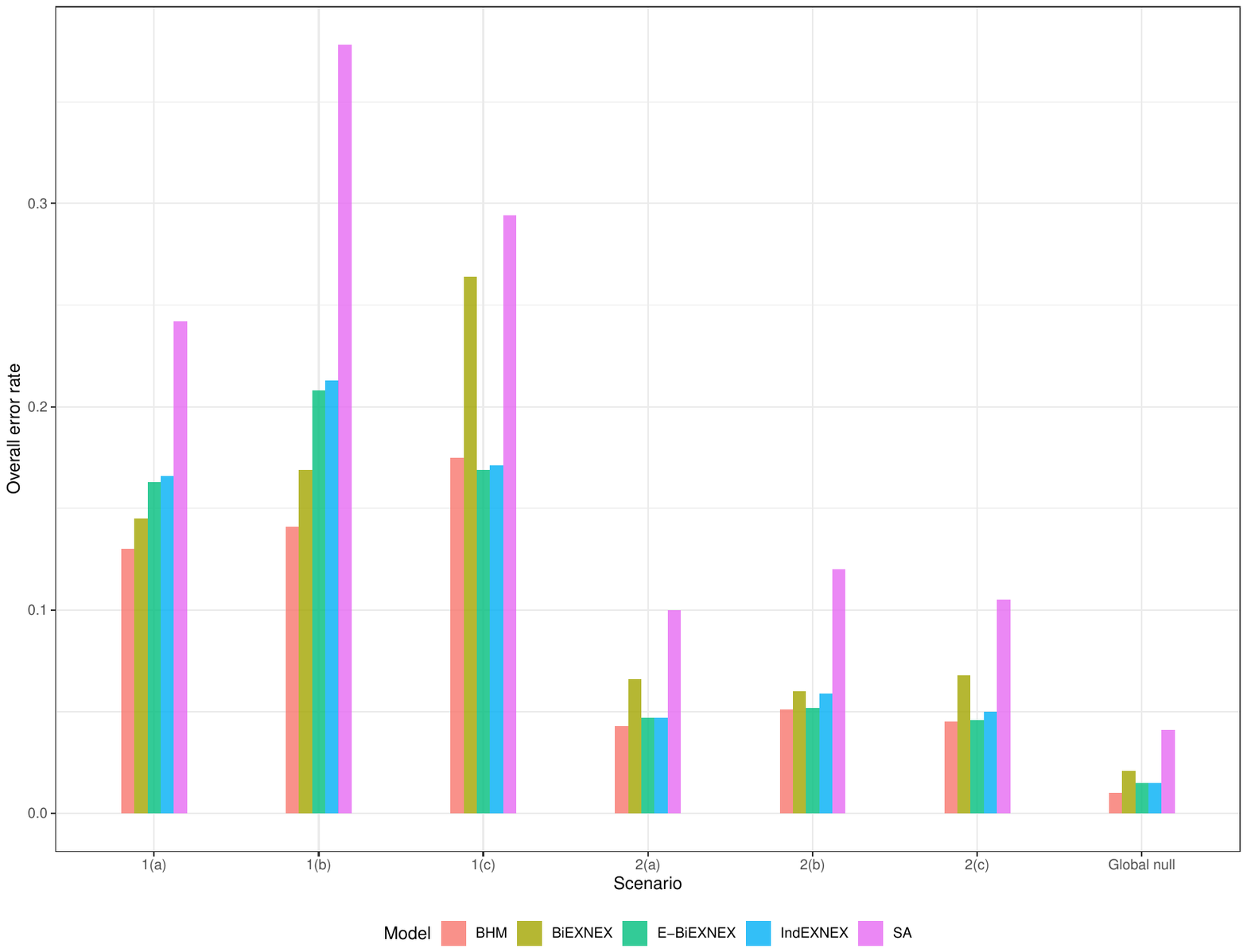}
	\end{center}
	\caption{The overall error rate under each null scenario for the second decision rule.}
	\label{fig:oer2}
\end{figure}

\clearpage
\section{Sensitivity analysis with smaller sample sizes} \label{app:smaller_samples}

In this section, our objective is to conduct a sensitive analysis to test how changing sample sizes can influence power calculation in presence of asynchronous exchangeability of treatment effects. We add more varieties as new sub-scenarios of original power scenario IIIa and IIIb with different degrees of similarity:
\begin{itemize}
	\item Sub-scenarios based on IIIa ($\theta_k^e = 1.0$):
	\begin{itemize}
		\item 1.1: $\bm{\theta}^t = (-0.69, -1.2, -1.24, -0.74, -1.04, -0.8)$ with SD$(\bm{\theta}^t) \approx 0.25$
		\item 1.2: $\bm{\theta}^t = (-0.97, -0.6, -1.09, -0.48, 0.38, -1.72)$ with SD$(\bm{\theta}^t) \approx 0.75$
		\item 1.3: $\bm{\theta}^t = (-3.82, -1.24, -0.4, 0.12, 0.11, 0.36)$ with SD$(\bm{\theta}^t) \approx 1.50$
		\item 1.3: $\bm{\theta}^t = (-2.46, 4.17, -0.28, -2.64, 0.79, -2.83)$ with SD$(\bm{\theta}^t) \approx 3.00$
	\end{itemize}
	\item Sub-scenarios based on IIIb ($\theta_k^t = -1.2$):
	\begin{itemize}
		\item 2.1: $\bm{\theta}^e = (0.77, 1.40, 1.15, 0.86, 1.25, 0.90)$ with SD$(\bm{\theta}^e) \approx 0.25$
		\item 2.2: $\bm{\theta}^e = (2.15, 0.36, 1.08, 1.86, 0.89, 0.47)$ with SD$(\bm{\theta}^e) \approx 0.75$
		\item 2.3: $\bm{\theta}^e = (2.86, 1.27, -1.51, -0.59, -0.68, 0.43)$ with SD$(\bm{\theta}^e) \approx 1.50$
		\item 2.4: $\bm{\theta}^e = (-1.26, 4.70, -3.55, 0.09, -3.18, 2.13)$ with SD$(\bm{\theta}^e) \approx 3.00$
	\end{itemize}
\end{itemize}

Each series of sub-scenarios under IIIa (or IIIb) have large to small similarities where the shared information in $\theta_k^t$ (or $\theta_k^e$) gradually diminishes. The (half-)sample sizes for subtrials are (5, 5, 5, 3, 6, 5), and all other parameters, priors and decision rules used for each method are identical as in Section \ref{sec:sim} to produce comparable results.
\begin{figure}[ht]
	\begin{center}
		\includegraphics[width=\textwidth]{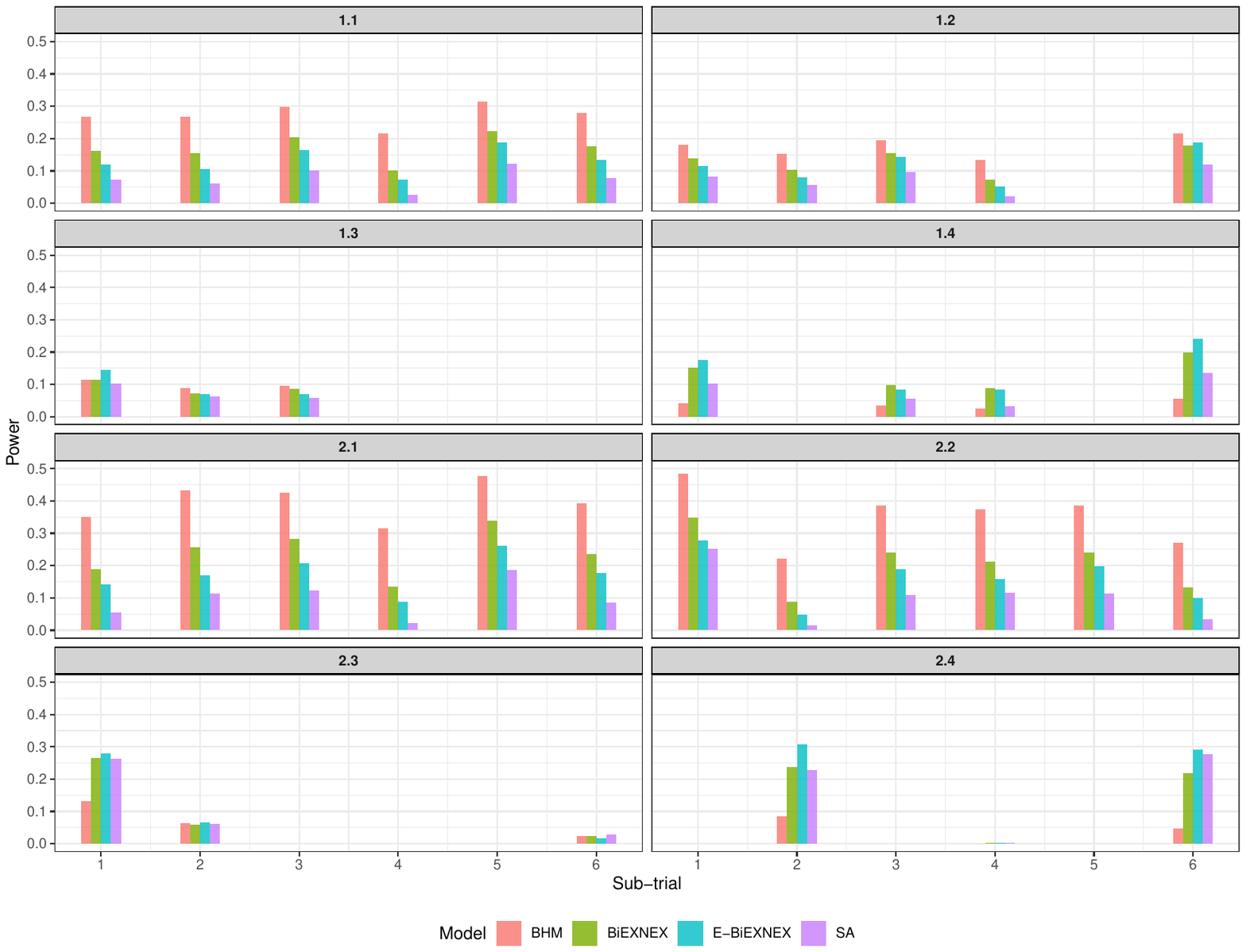}
	\end{center}
	\caption{Bayesian analogues of power for the smaller sample sizes.}
	\label{fig:small_sample_appendix}
\end{figure}

Figure \ref{fig:small_sample_appendix} shows that regardless of which treatment effect is the same across all subtrials, BHM will achieve lower power when the similarity of another treatment effect decreases while E-BiEXNEX can gradually strike the highest power among all methods (e.g. from 1.1 to 1.4 or 2.1 to 2.4). In contrast to the scenario IIIb in Section \ref{sec:sim_results} where $\theta_k^t$ are the same for all subtrials and $\theta_k^e$ are largely different, sub-scenarios 2.3 and 2.4 in this section present a reverse pattern, proving that the smaller sample sizes cannot provide sufficient information even for continuous outcomes. Hence, borrowing of information depends on both treatment effects.

\clearpage
\section{Simulation results under the first decision rule} \label{app:sim_results}

\begin{figure}[ht]
	\begin{center}
		\includegraphics[width=0.9\textwidth]{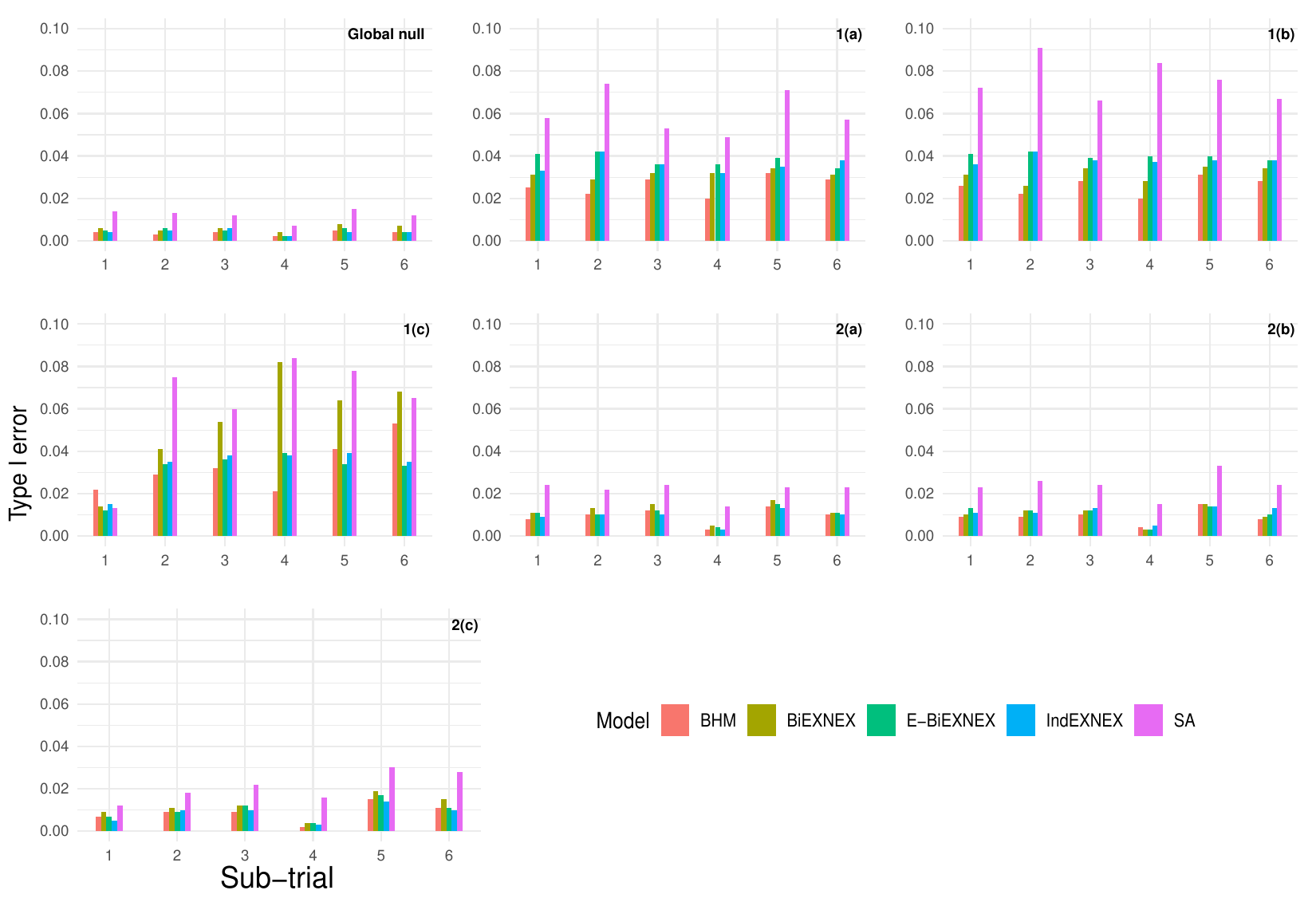}
	\end{center}
	\caption{Bayesian analogues of type I error rates for each null scenario in Section \ref{sec:dmrule}.}
	\label{fig:typeIerr_appendix}
\end{figure}

\text{ }
\begin{figure}[ht]
	\begin{center}
		\includegraphics[width=0.9\textwidth]{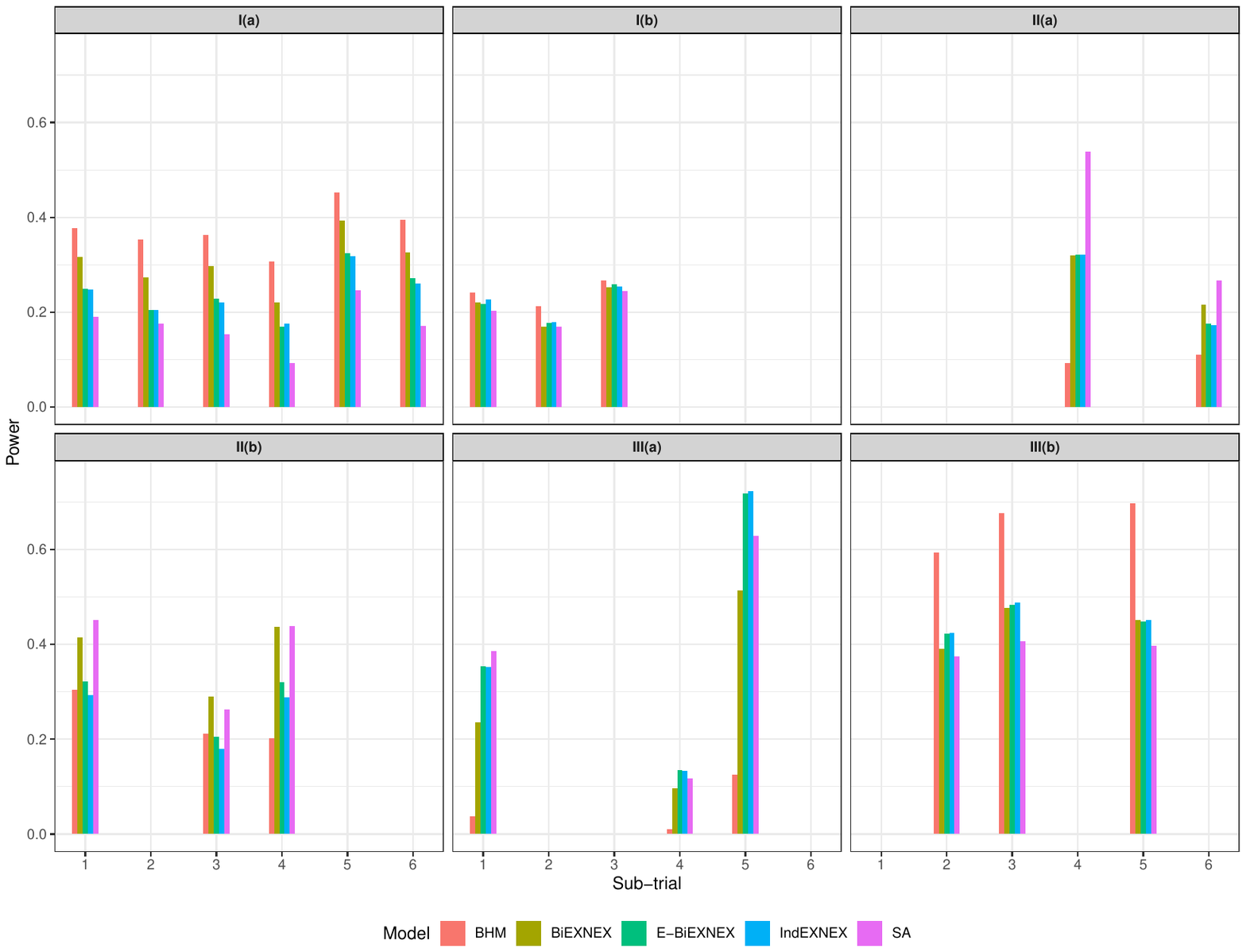}
	\end{center}
	\caption{Bayesian analogues of power for each scenario, model, and subtrial.}
	\label{fig:power_appendix}
\end{figure}

\clearpage

\bibliographystyle{apalike}
\bibliography{ref}

\clearpage

\renewcommand\thefigure{\arabic{figure}}    
\setcounter{figure}{0}


\end{document}